\documentclass[a4paper,11pt]{article}
\pdfoutput=1
\usepackage{jcappub}
\usepackage{booktabs}
\usepackage[T1]{fontenc}
\usepackage[utf8]{inputenc}
\usepackage{txfonts}
\usepackage{url}
\usepackage{jcappub} 
\usepackage{appendix}
\usepackage{graphicx}
\usepackage{dcolumn}
\usepackage{amssymb}
\usepackage{amsfonts}
\usepackage{amsbsy}
\usepackage{color}
\usepackage{rotating}
\usepackage[none]{hyphenat}
\usepackage[english]{babel}
\renewcommand{\H}{{\cal H}}

\title{Linear and non-linear perturbations in dark energy models}

\author[b,a]{Celia Escamilla-Rivera,}
\author[b]{Luciano Casarini,}
\author[b,d]{J\'ulio C. Fabris}
\author[c]{and Jailson S. Alcaniz}

\affiliation[b]{Mesoamerican Centre for Theoretical Physics, Universidad Aut\'onoma de Chiapas, 29040, Tuxtla Guti\'errez, Chiapas, M\'exico.}
\affiliation[a]{Departamento de Fisica, Universidade Federal do 
Espirito Santo, Av. Fernando Ferrari 514, Vitoria, ES, 29075-910, 
Brasil}
\affiliation[c]{Observat\'orio Nacional, 20921-400, Rio de Janeiro - RJ, Brasil}
\affiliation[d]{National Research Nuclear University ``MEPhl'', Kashirskoe sh. 31, Moscow 115409, Russia}

\emailAdd{cescamilla@mctp.mx, casarini.astro@gmail.com, julio.fabris@cosmo-ufes.org, alcaniz@on.br}

\abstract 
{In this work we discuss observational aspects of three time-dependent  
parameterisations of the dark energy equation of state $w(z)$. In order to determine the dynamics associated with these models, we calculate their background evolution and perturbations in a  scalar field representation. After performing a complete treatment of linear perturbations, we also show that the non-linear contribution of the selected $w(z)$ parameterisations to the matter power spectra is almost the same for all scales, with no significant difference from the predictions of the standard $\Lambda$CDM model. }

\begin{document}
\maketitle
\flushbottom

\section{Introduction}

Probing the nature of the physical mechanism behind the current 
cosmic acceleration is one of the central issues in theoretical 
physics and cosmology. In the framework of the standard $\Lambda$CDM 
model, which seems to be consistent with most of the cosmological 
observations, the observed acceleration is explained by adding a 
cosmological constant $\Lambda$ to the right-hand side of Einstein 
field equations.  Despite of its simplicity and success in 
explaining present-day data, the standard cosmology has a couple of 
theoretical loopholes as, for example, the fine tuning and 
coincidence problems~\cite{Weinberg:2000yb}, which have led to 
alternative proposals that either modify the Einstein field equations on large 
scales or consider a landscape with a dynamic dark energy (we refer 
the reader to~\cite{Sahni:1999gb, Padmanabhan:2002ji, 
Peebles:2002gy, Copeland:2006wr} for some reviews).

Following the latter approach, the dark energy component is 
described by an equation-of-state (EoS) parameter, $w(z)$, which 
evolves with the redshift, being  physically restricted to the 
interval $-1 \leq w(z) \leq -1/3$ (for a discussion on the so-called 
phantom fields, for which $w < -1$, see, e.g.,~\cite{Caldwell:1999ew, 
Carroll:2003st, Alcaniz:2003qy, Lima:2004wf, Lima:2003dd}). Currently,  there is 
no strong observational evidence either for departures from $w = -1$ 
or for a time evolution of the dark energy EoS. However, since such 
results would be of great impact on cosmology, a number of studies 
on dark energy parameterisations have been discussed in the 
literature (see, e.g., \cite
{Feng:2011zzo,Stefancic:2005sp,Jassal:2004ej,Wang:2008zh,Wang:2005yaa,
Barboza:2009ks} and references therein).

Observational constraints on time-dependent EoS parameterisations 
have been obtained using different observables, such as distance 
measurements to type Ia supernovae (SNe Ia)~\cite{Betoule:2014frx, santos2016}, 
measurements of the baryonic acoustic oscillation (BAO) scale \cite
{Busca:2012bu}, anisotropies of the cosmic microwave background 
(CMB) \cite{Ade:2015xua}, among others~\cite
{Hicken:2009dk,Stern:2009ep}. Presently, these observations allow 
for slight deviations from the standard model ($w = -1$), which are 
usually characterised by two parameters ($w_0$, $w_a$). 
 
The goal of the present analysis is to investigate the non-linear 
contribution of some selected dark energy parameterisations to the 
matter power spectra and use this observable to infer a possible 
time-dependence of $w$. In our analysis, we consider three EoS 
parameterisations, as discussed in Refs.~\cite{cooray, weller, cp, 
linder, Barboza:2008rh}. After fitting their parameters to the 
current SNe Ia and BAO datasets, a complete treatment of the linear 
evolution of perturbations from the entry of perturbations produced 
by inflation in the horizon until today is presented. Firstly, we solve the perturbation equations for different modes  using a  scalar field representation. Then, we modify the code CAMB \cite{camb} by implementing the Parameterised 
Post-Friedmann (PPF) \cite{PPF} approach to cross the phantom 
divide line, which allows to study the linear matter power spectrum. In 
addition, we also estimate the non-linear matter power spectrum by extending 
the HALOFIT \cite{halofit1,halofit2} routine, built originally for $w=const.$ 
models, to the time-varying equation-of-state parameterisations considered in this work. For this purpose, we 
use a suitable spectral equivalence described in \cite
{2009JCAP...03..014C, pkequal}.

The structure of this paper is the following: In Section \ref 
{sec:backeqs} we review the context of the background equations for 
the scalar field dynamics. We discuss the dark energy parameterisations in a scalar field representation in Section \ref{sec:DE_Parameterisations}. In Section \ref{sec:observations} we review the current constraints from type Ia supernovae and BAO observations on these parameterisations and present a brief 
comparison with the $\Lambda$CDM model,  showing, in particular, that a possible \textit{tension} between them is minimal \cite{EscamillaRivera:2011qb}. In 
Section \ref{sec:linear_perturbations} we present the linear 
perturbations of the dark energy parameterisations coupled to a 
scalar field, and show that the solution is 
scale invariant.
The non-linear contribution of the selected dark energy parameterisations to the 
matter power spectra is discussed in Section \ref {sec:non_linear}.  
Finally, in Section \ref{sec:conclusions}, we 
summarise our main conclusions and results.

\section{Background equations}
\label{sec:backeqs}

Following standard lines, we consider that dark energy and matter 
(baryonic + dark) exchange preserves, separately for each component, the total energy conservation 
equation 
\begin{eqnarray}
\dot{\rho} +3\frac{\dot{a}}{a} (\rho+p)=0, 
\end{eqnarray}
or still 
\begin{eqnarray}
\frac{d\rho}{\rho}=-3 \frac{da}{a}(1+w)\;,
\end{eqnarray}
where the EoS parameter $w=p/\rho$ is the ratio between the pressure 
and the energy density. 

\subsection{Scalar field dynamics}

The scalar field representation of dark energy parameterisations 
can be done by considering the equations for density and pressure of 
a scalar field $\psi$ as follows~\cite{Barboza:2011py}

\begin{eqnarray}
\rho_{\psi} &=& \frac{\dot{\psi}^2}{2} +U,  \\
p_{\psi} &=& \frac{\dot{\psi}^2}{2} -U, 
\end{eqnarray}
from where we can rewrite 
\begin{eqnarray}
\dot{\psi}^2 &=& (w+1) \rho_{\psi}, \\
U&=& \frac{\rho_{\psi}}{2}(1-w).
\end{eqnarray}

Using this relations with $8 \pi G \rho_{\psi,0} = 3 H_{0}^{2} \Omega_{\psi}$ we can 
obtain a set of these functions for any specific form of $w(z)$. As 
we will discuss later, with these expressions it is possible to 
compute the $w(z)$ contribution to the linear and non-linear 
perturbations. 
The expansion rate is related to the energy content with the Friedmann 
equation, as usual:
\begin{eqnarray}
H^{2} = H_0^2 \left[ \Omega_m (1+z)^3 + \Omega_{\psi} e^{3\int^{z}_{0}{\frac{1+w(z')}{1+z'}}dz'} \right ].
\end{eqnarray}

\section{Parameterisations and their scalar field representation}
\label{sec:DE_Parameterisations}

Taylor series-like parameterisations of the type $w(z) = \sum_{n = 
0}{w_nx_n(z)}$, where $w_n$ are constants and $x_n(z)$ are functions 
of the scalar factor, $a$, or, equivalently, the redshift $z$, are 
among the most commonly adopted in the literature. However, their 
analysis does not take into account a canonical scalar field, which 
can be a good candidate for the observed cosmic expansion. In this work we 
consider three $w(z)$ parameterisations and their scalar field representations. 

\subsection{Linear Model}

The simplest way to parameterise the evolution of the equation of state 
$w$ is by taking a Taylor expansion at first-order~\cite{cooray, weller}
\begin{eqnarray} \label{p1}
w(a)&=&w_0 +w_1 \frac{(a-1)}{a}, \quad \text{or} \quad w(z)=w_0 -w_1 z,
\end{eqnarray}
which can be reduced to $\Lambda$CDM model ($w(z)=w=-1$) for $w_0 = 
-1$ and $w_1 =0$. The energy density associated to this model is 
then given by
\begin{eqnarray}
\rho_{\psi} &=& \rho_{\psi,0} e^{3w_1 \left(1-\frac{1}{a}\right)} a^{-3(w_0 +w_1 +1)},
\end{eqnarray}
where we consider a normalisation with $a_0 =1$. Using the set of 
background equations given above we can calculate the scalar field and 
the potential for this model
\begin{eqnarray}
8 \pi G \dot{\psi}^2 &=&\left( 1+w_0 +w_1 -\frac{w_1}{a} \right) 3H_0^2\Omega_{\psi}
e^{3 w_1 \left(1-\frac{1}{a}\right)}a^{-3(1+w_0 
+w_1)}, \quad\quad \\
8 \pi G U &=& \left(1-w_0 -w_1 +\frac{w_1}{a}\right)  \frac{3}{2} H_0^2 \Omega_{\psi} 
e^{3w_1 \left(1-\frac{1}{a}\right)} a^{-3(1+w_0 +w_1)}.
\end{eqnarray}
The Hubble expansion rate can be written as
\begin{eqnarray}
H=H_0\sqrt{\Omega_m a^{-3}+\Omega_{\psi} a^{-3(1+w_0 +w_1)}e^{3w_1 
(1-\frac{1}{a})}}.
\end{eqnarray}

\subsection{Chevallier-Polarski-Linder (CPL) model}

Currently, the most adopted parameterisation is the so-called 
Chevallier-Polarski-Linder (CPL) parameterisation~\cite{cp, linder}
\begin{eqnarray}
w(a)&=& w_0 +w_1 (1-a) \quad \text{or} \quad
w(z)=w_0 +\frac{z}{1+z} w_1. \quad\quad
\end{eqnarray}
The energy density associated to this model is given by
\begin{eqnarray}
\rho_{\psi} &=& \rho_{\psi,0} a^{-3(1+w_0 +w_1)} e^{3w_1 (a-1)}.
\end{eqnarray}
Note that, differently from (\ref{p1}), the CPL parameterisation 
does not blow up as $e^{3w_1 (1-\frac{1}{a})}$ in the past. On the 
other hand, it does blow up exponentially in the future as $a 
\rightarrow \infty$ ($z \rightarrow -1$) for $w_1 >0$~\cite
{Wang:2004py}.

Using the background equations we can calculate the scalar field and 
the potential for this model
\begin{eqnarray}
8 \pi G \dot{\psi}^2 &=& (1+w_0+w_1 
-aw_1 ) 3 H_{0}^2 \Omega_{\psi} a^{-3(1+w_0 +w_1)} e^{3 w_1 
(a-1)}, \quad\quad  \\
8 \pi G U &=& \left(1-w_0 -w_1 +aw_1\right)\frac{3}{2} H_0^2 \Omega_{\psi} a^{-3(1+w_0 
+w_1)} e^{3w_1 (a-1)}.
\end{eqnarray}
For this model, the Hubble expansion rate is given by
\begin{eqnarray}
H=H_0\sqrt{\Omega_m a^{-3}+\Omega_{\psi} a^{-3(1+w_0 +w_1)}e^{3w_1(a-1)}}.
\end{eqnarray}

\subsection{Barboza-Alcaniz (BA) model}

This model, proposed in \cite{Barboza:2008rh}, is well-behaved over 
the entire cosmic evolution and mimics a linear-redshift evolution 
at low redshift. Its functional form is given by
\begin{eqnarray}
w(a)&=& w_0 +w_1 \left(\frac{1-a}{2a^2 -2a +1}\right)\quad 
\text{or}\quad
w(z)=w_0 +w_1 \frac{z(1+z)}{1+z^2}. \quad\quad
\end{eqnarray}
Note that this parameterisation does not diverge at $z\rightarrow 
-1$ as the above ones. The energy density associated to this model is 
given by
\begin{eqnarray}
\rho_{\psi} &=&  \rho_{\psi,0} a^{-3(1+w_0+w_1)} \left[1+2(a-1)a\right]^{3w_1 /2}.
\end{eqnarray}

Using the background equations we can calculate the scalar field and 
the potential for this parameterisation, i.e.,
\begin{eqnarray}
8 \pi G \dot{\psi}^2 &=& \left[ 1+w_0 +w_1 \left(\frac{1-a}{2a^2 -2a 
+1}\right) \right] 3 H_0^2 \Omega_{\psi} a^{-3(1+w_0 +w_1)} \times 
\left[1+2(a-1)a\right]^{\frac{3}{2}w_1}, \\
8 \pi G U &=&  \left[ 1-w_0 -w_1 \left(\frac{1-a}{2a^2 -2a 
+1}\right) \right] \frac{3}{2} H_0^2 \Omega_{\psi} a^{-3(1+w_0 +w_1)} \times 
\left[1+2(a-1)a\right]^{\frac{3}{2}w_1}.
\end{eqnarray}

The Hubble expansion rate can  be written as
\begin{eqnarray}
H=H_0\sqrt{\Omega_m a^{-3}+\Omega_{\psi} a^{-3(1+w_0 +w_1)}e^{-3w_1 
\left(\frac{1-a}{2a^2 -2a +1}\right)}}. \quad
\end{eqnarray}

\begin{figure}[tbp]
\centering 
\includegraphics[width=0.8\textwidth,origin=c,angle=0]{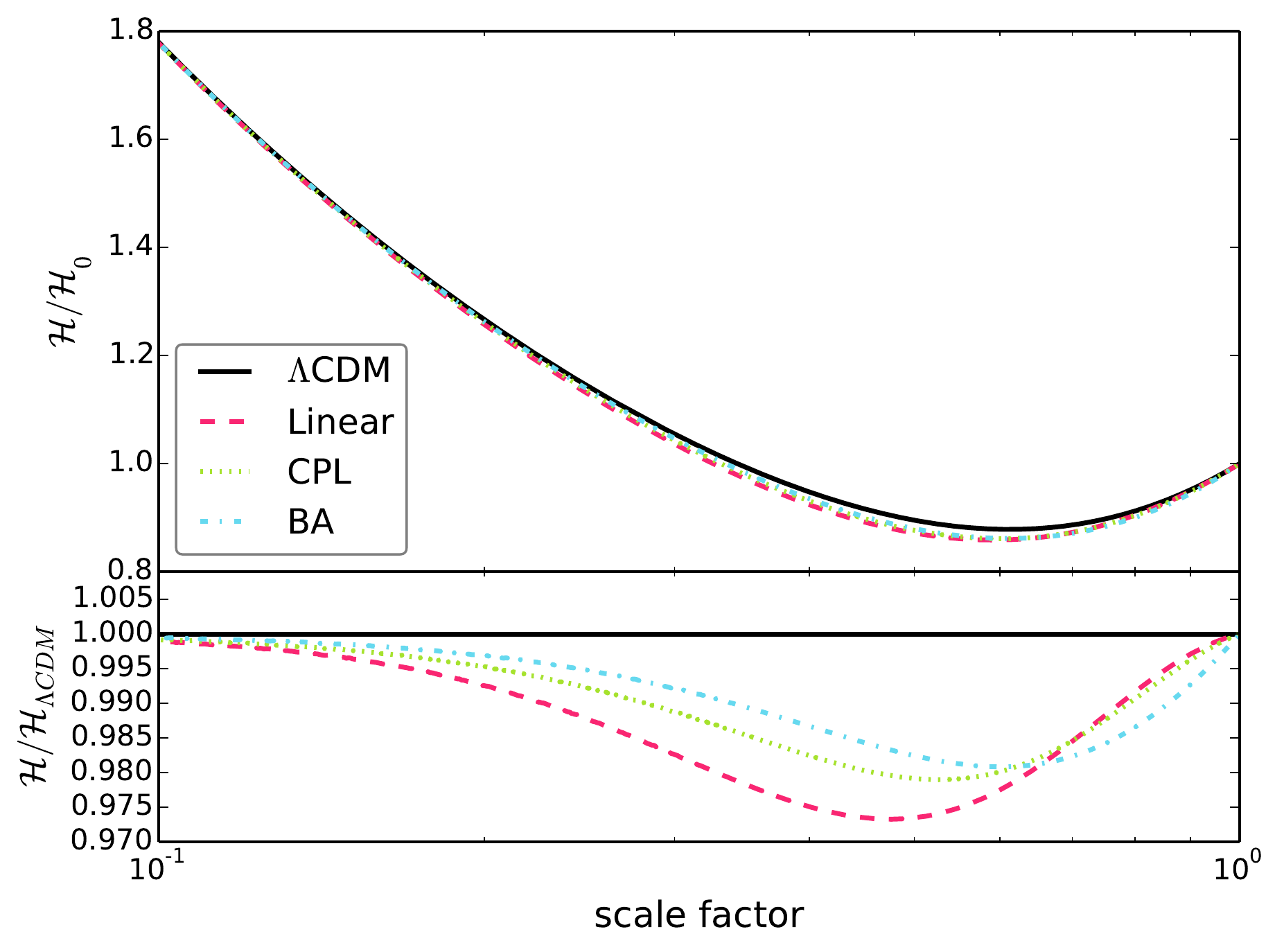}
\caption{\label{fig:cosmo_evolution} Evolution of the conformal Hubble 
parameter ($\H=aH$) for the three dark energy parameterisations discussed in the text. The curves are obtained for the best-fit values given by the joint JLA  +
BAO analysis.} \end{figure}

The evolution of these dark energy models in comparison with 
$\Lambda$CDM are displayed in Figure \ref{fig:cosmo_evolution} using 
the datasets described in Sec. \ref{sec:observations}.

\section{Current observational constraints}
\label{sec:observations}

{\renewcommand{\tabcolsep}{10.mm}
{\renewcommand{\arraystretch}{1.4}
\begin{table}
\caption{JLA supernovae binned sample 
data}\label{tab:dataJLA}
\centering
\resizebox*{0.5\textwidth}{!}{
\begin{tabular}{ccccc}
\hline
\hline
{Redshift}  &  {$\mu$}  & {${\sigma_{\mu}}^2$} \\
\hline
\hline
$0.01$   & $32.954$   & $0.021$  \\
$0.012$  & $33.879$   & $0.028$  \\
$0.014$  & $33.842$   & $0.006$ \\
$0.016$  & $34.119$   & $0.005$  \\
$0.019$  & $34.593$   & $0.007$ \\
$0.023$  & $34.939$   & $0.003$  \\
$0.026$  & $35.252$   & $0.004$  \\
$0.031$  & $35.749$   & $0.003$  \\
$0.037$  & $36.069$   & $0.003$ \\
$0.043$  & $36.436$    & $0.006$  \\
$0.051$  & $36.651$    & $0.009$  \\
$0.060$  & $37.158$    & $0.004$ \\
$0.070$  & $37.430$    & $0.004$  \\
$0.082$  & $37.957$     & $0.003$  \\
$0.097$  & $38.253$     & $0.004$  \\
$0.114$  & $38.613$     & $0.001$  \\
$0.134$  & $39.068$     & $0.001$  \\
$0.158$  & $39.341$     & $0.001$  \\
$0.186$  & $39.792$     & $0.001$ \\
$0.218$  & $40.157$     & $0.001$  \\
$0.257$  & $40.565$     & $0.001$  \\
$0.302$  & $40.905$     & $0.002$  \\
$0.355$  & $41.421$     & $0.001$  \\
$0.418$  & $41.791$     & $0.001$  \\
$0.491$  & $42.231$     & $0.002$  \\
$0.578$  & $42.617$     & $0.001$  \\
$0.679$  & $43.053$     & $0.004$  \\
$0.799$  & $43.504$     & $0.003$  \\
$0.940$  & $43.973$     & $0.004$  \\
$1.105$  & $44.514$     & $0.024$  \\
$1.3$  & $44.822$     & $0.019$  \\
\hline
\hline
\end{tabular}}
\end{table}}}

\subsection{SNe Ia}
For the first cosmological test we will employ the most recent SNe 
Ia catalog available, the JLA (acronym for Joint Lightcurve 
Analysis) described in \cite{Betoule:2014frx}. Given the same trend 
as using the full catalog itself, we employ here the catalog with a 
binned sample data described in Table \ref{tab:dataJLA} which 
consists of $N_{\text{JLA}}=31$ SNe Ia events distributed over the 
redshift interval $0.01< z <1.3$. The statistical analysis of the 
this binned data lies in the definition of the modulus distance: 
\begin{eqnarray}
\mu(z_i, \mu_0) = 5 \log_{10}\left[d_L (z_i, 
\Omega_m;w_0,w_1)\right] +\mu_0,
\end{eqnarray}
where
\begin{eqnarray}
d_L(z,\Omega_m;w_0,w_1) = 
(1+z)\int_{0}^{z}{d\tilde{z}E^{-1}(\tilde{z},\Omega_m;w_0,w_1)},
\end{eqnarray}
is the Hubble free luminosity distance with $E=H(z)/H_0$ and ($w_0$,
$w_1$) are the free parameters of the model. The best fit values are 
obtained by minimizing the quantity
\begin{eqnarray} 
\chi_{\text{SN}_{\text{JLA}}}^2 
=\sum^{N_{\text{JLA}}}_{i=1}{\frac{\left[\mu(z_i ,\Omega_m ;\mu_0, 
w_0,w_1)-\mu_{\text{obs}}(z_i)\right]^2}{\sigma^{2}_{\mu,i}}},
\end{eqnarray}
where the $\sigma^{2}_{\mu,i}$ are the measurements errors.

\subsection{Baryon acoustic oscillations}
 
The sample of BAO measurements used in this analysis 
is described in \cite{Beutler:2011hx,Xu:2012hg,Anderson:2013zyy}.
Before proceeding to the statistical analysis of these data, we 
define the ratio 
\begin{equation}
d_{z} \equiv \frac{r_{s}(z_{d})}{D_{V}(z)},
\end{equation}
where $r_{s}(z_{d})$ is the comoving sound horizon at the drag epoch
\begin{equation}
r_{s}(z_{d}) = \frac{c}{H_{0}} \int_{z_{d}}^{\infty} 
\frac{c_{s}(z)}{E(z)} \mathrm{d}z\; ,
\end{equation}
with $c$ being the light velocity, $c_{s}$ the sound speed and 
$z_{d}$ the redshift of the drag epoch. By definition the dilation scale 
$D_{V}(z)$ is 
\begin{equation}
D_{V}(z,\Omega_m; w_0,w_1) = \left[ (1+z)^2 D_{A}^2 \frac{c 
\, z}{H(z, \Omega_m; w_0,w_1)} \right]^{1/3},
\end{equation}
where $D_{A}$ is the angular diameter distance:
\begin{equation}
D_{A}(z,\Omega_m; w_0,w_1) = \frac{1}{1+z} \int_{0}^{z} 
\frac{c \, \mathrm{d}z'}{H(z', \Omega_m;w_0,w_1)} \; .
\end{equation}

Through the comoving sound horizon, the distance ratio $d_{z}$ is 
related to the expansion parameter $h$ (defined such that $H \doteq 100 
h$ km/s/Mpc) and the physical densities $\Omega_{m}$ and $\Omega_{b}$. 
Specifically, we have
\begin{equation}
r_{s}(z_{d}) = 153.5 \left( \frac{\Omega_{b} 
h^2}{0.02273}\right)^{-0.134} \left( \frac{\Omega_{m} 
h^2}{0.1326}\right)^{-0.255} \; \mathrm{Mpc}, \; 
\end{equation}
with $\Omega_b=0.045\pm0.00054$.

The $\chi^2$ function for the BAO data is defined as:
\begin{equation}\label{chibao}
\chi^2_{\mathrm{BAO}}(\boldsymbol{\theta}) = 
\mathbf{X}^T_{\mathbf{BAO}}
\mathbf{C}^{-1}_{\mathbf{BAO}} 
\mathbf{X}_{\mathbf{BAO}},
\end{equation}
where $\mathbf{X}_{\mathbf{BAO}}$ is given as
\begin{equation}
\mathbf{X_{BAO}}=\left(\begin{array}{c}
\frac{r_s (z_d)}{D_V (0.106,\Omega_m;w_0,w_1})    -0.336\\
\frac{r_s (z_d)}{D_V (0.35,\Omega_m;w_0,w_1})    -0.1126 \\
\frac{r_s (z_d)}{D_V (0.57,\Omega_m;w_0,w_1})    -0.07315\\
\end{array} \right)\;,
\end{equation}
and
\begin{eqnarray}
\mathbf{C}^{-1}_{\textbf{BAO}}=\text{diag}(4444,215156,721487),
\end{eqnarray}
In order to determine the best fit values of the parameters $w_0$ 
and $w_1$ for our three parameterisations discussed above, we will 
employ the maximum likelihood method, where the total likelihood for 
joint data analysis is expressed as the sum of each dataset, i.e.,
\begin{equation}
\chi_{\text{Total}}^2 =\chi_{\text{SN}_{\text{JLA}}}^2 
+\chi^2_{\text{BAO}}.
\end{equation}

\begin{figure}[tbp]
\centering 
\includegraphics[width=0.49\textwidth,origin=c,angle=0]{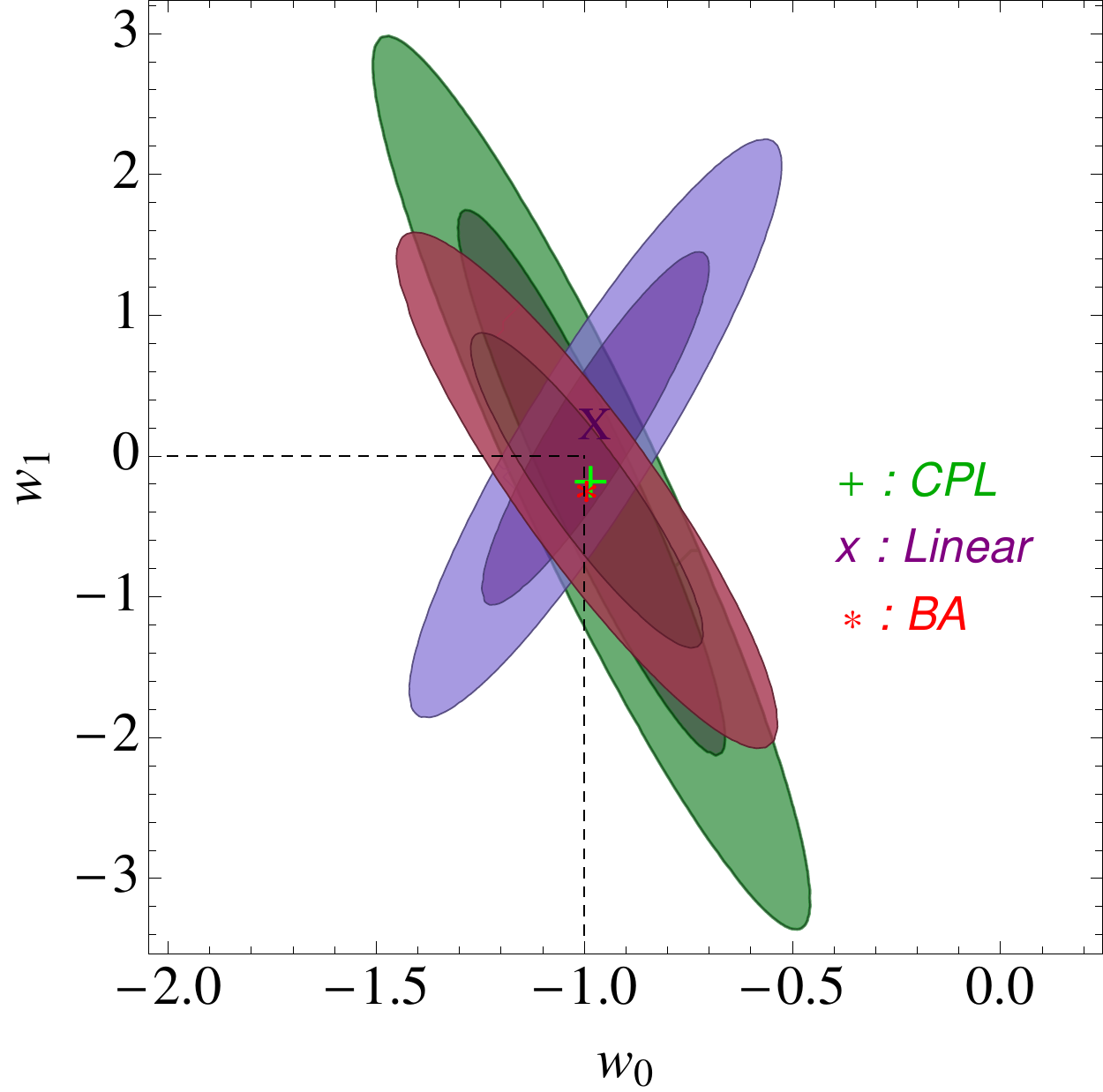}
\caption{\label{fig:cosmo_test2} 
Comparison between three dark energy 
parameterisations at 1$\sigma$ and 2$\sigma$ confidence contours tested with 
JLA supernovae binned data sample. The dashed line represent the 
$\Lambda$CDM model. The best fit points for each model are represented by a `green plus sign'
in the case of the CPL model, a `purple x sign' in the case of the Linear model and `red star sign' 
for the BA model \cite{Escamilla-Rivera:2016qwv}.
}
\end{figure}
\begin{figure}[tbp]
\centering 
\includegraphics[width=0.49\textwidth,origin=c,angle=0]{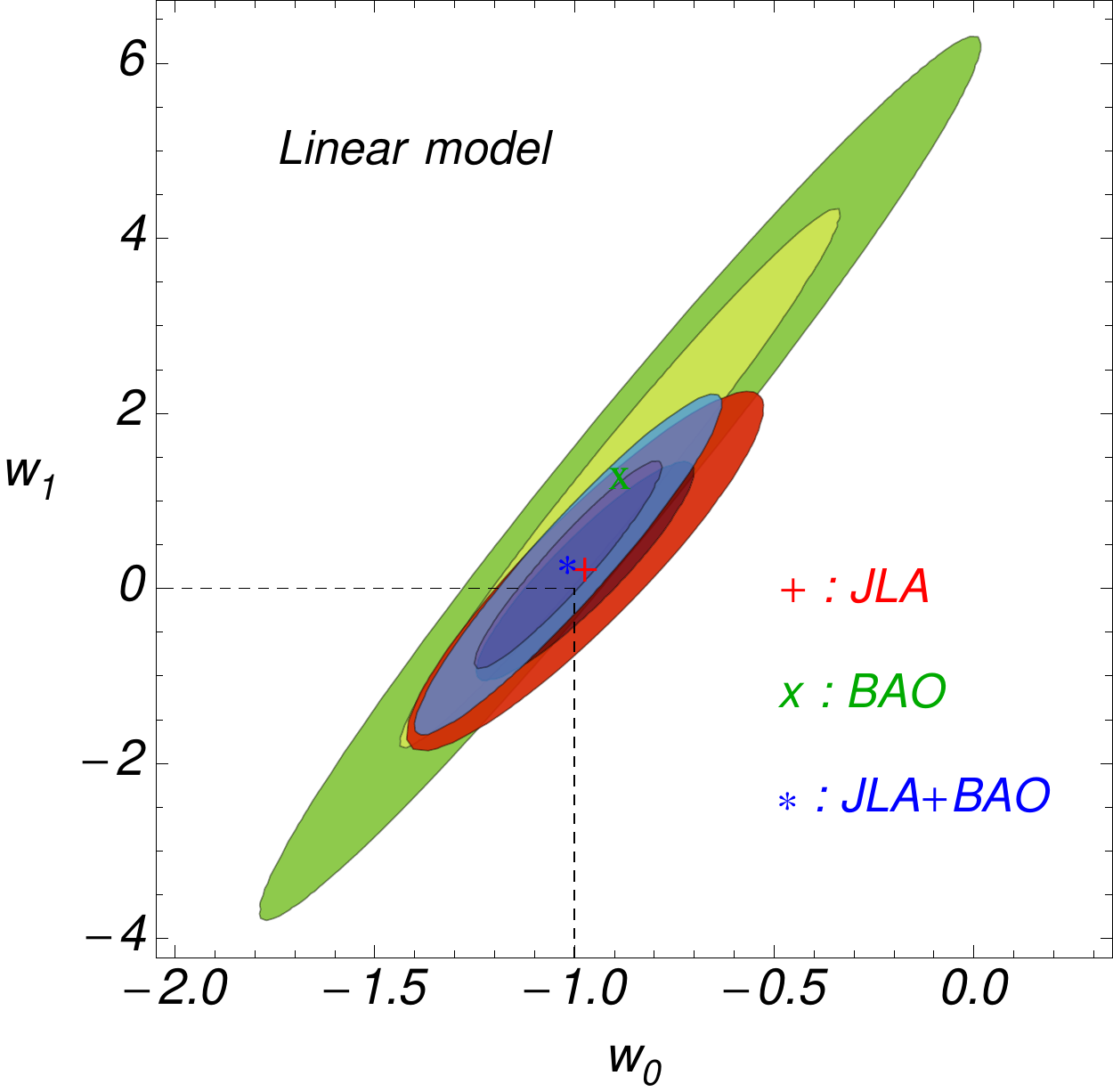}
\includegraphics[width=0.49\textwidth,origin=c,angle=0]{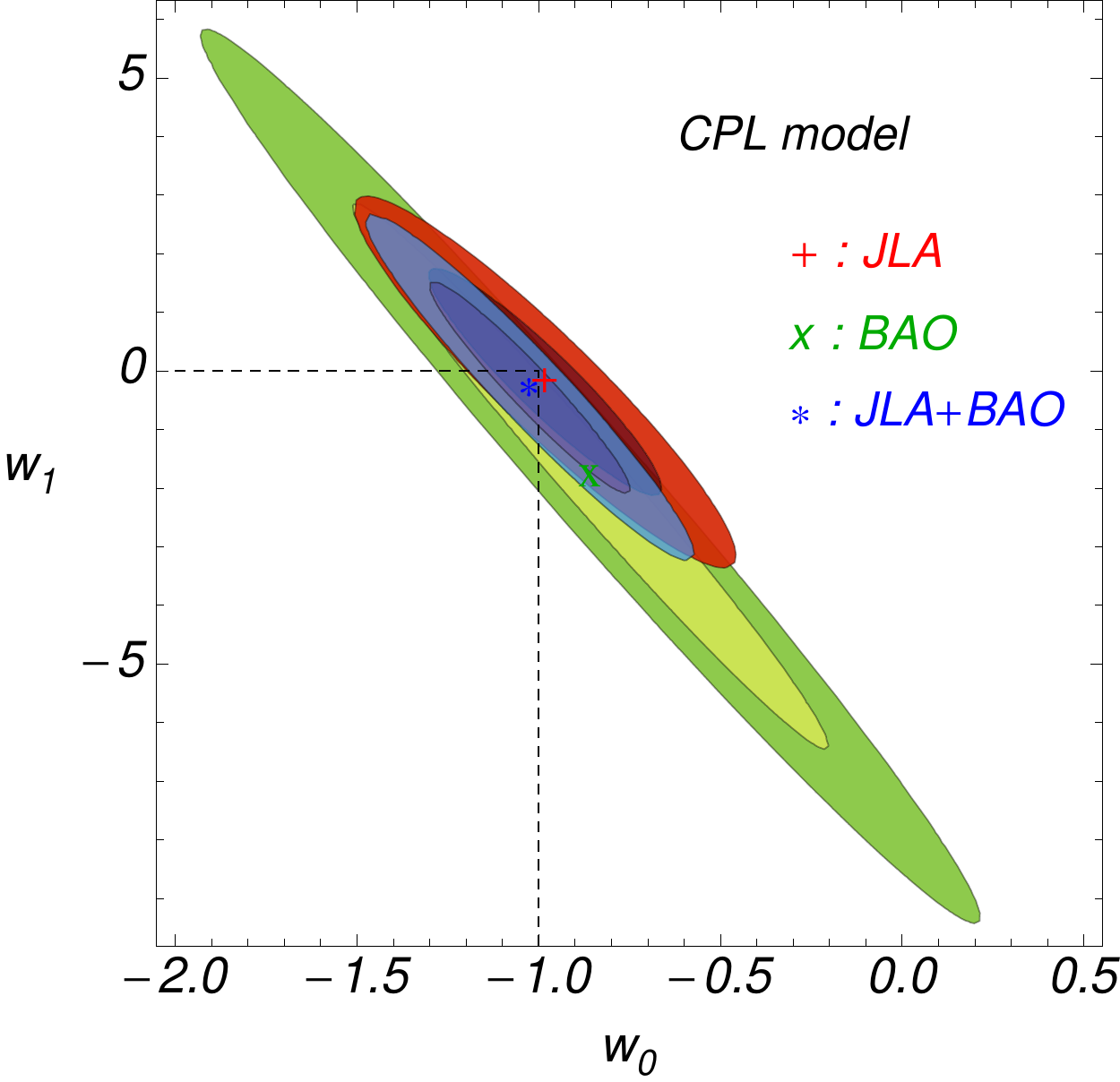}
\includegraphics[width=0.49\textwidth,origin=c,angle=0]{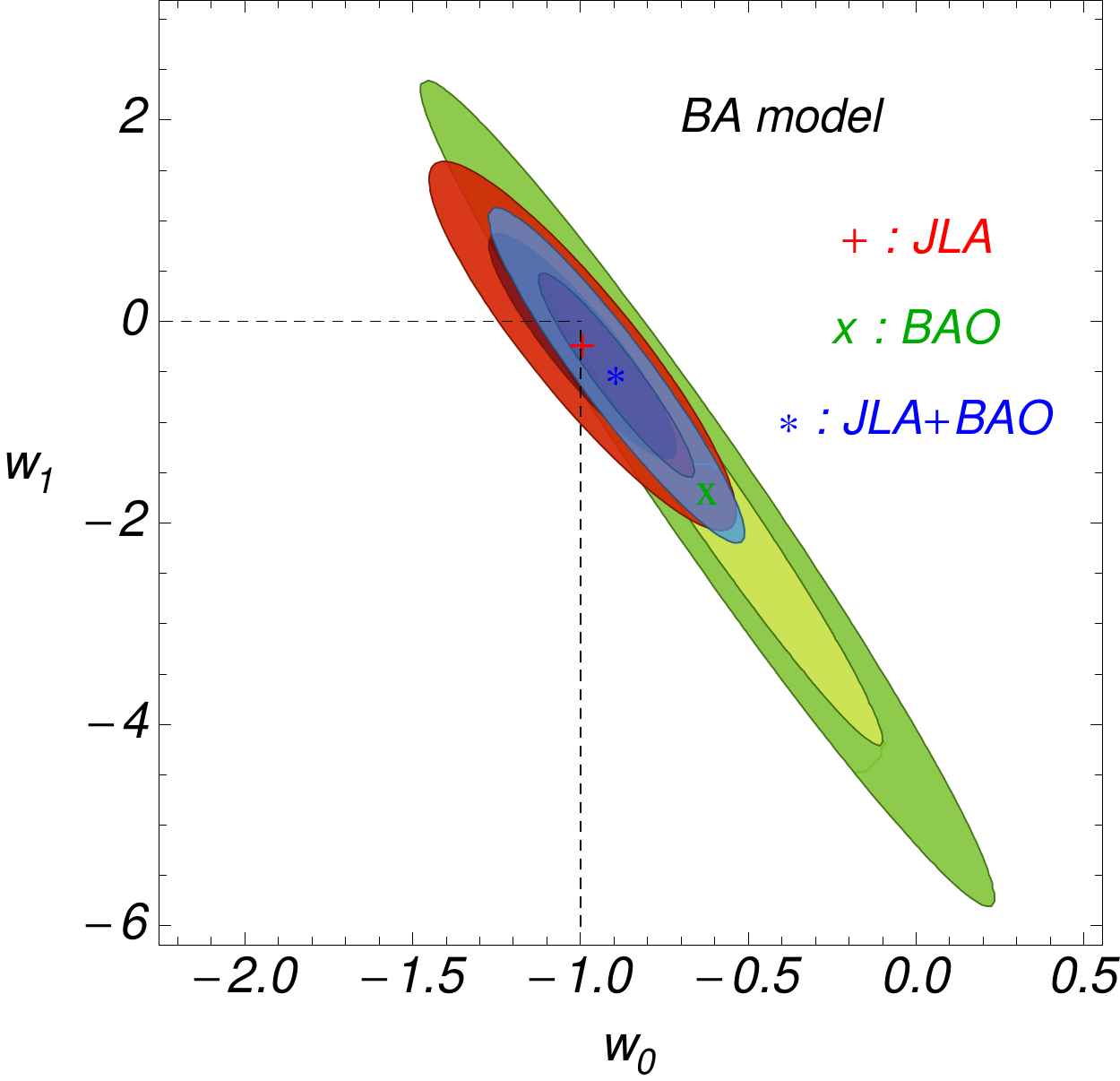}
\caption{\label{fig:cosmo_test4} 
1 and 2$\sigma$ confidence contours 
for the parameterisations discussed in the text. Constraints from the JLA supernovae binned data are 
represented by the red region whereas the BAO high-$z$ sample with 
($z=0.106,0.35,0.57$) are the green region. The combined 
JLA+BAO bounds are represented by the blue region. The point where the 
dashed lines cross indicates the $\Lambda$CDM model.
The best fit points for each dataset are represented by a `red plus sign'
in the case of the JLA sample, a `green x sign' in the case of the BAO sample and `blue star sign' 
for the combined JLA+BAO sample.
}
\end{figure}

{\renewcommand{\tabcolsep}{0.4mm}
{\renewcommand{\arraystretch}{2.}
\begin{table*}
\begin{minipage}{\textwidth}
\caption{Parameterisation models and data. Column.~1: Models; 
columns~2~-~3~-~4: best fit values of $w_0$ and $w_{1}$ using JLA 
supernovae binned and BAO samples; column. 5: $d_\sigma$ with 
respect to the $\Lambda$CDM model using (\ref{eq:dsigma}).}
\resizebox*{\textwidth}{!}{
\begin{tabular}{cccccccc}
\hline
\multicolumn{1}{c}{$\mathrm{Model}$} & \multicolumn{3}{c}{$\mathrm{Best \; fit \; parameters \; (w_{0}, w_{1})}$} & \multicolumn{3}{c}{$d_{\sigma}^{\mathrm{Model-\Lambda}}$}\\
\hline \hline
$ $ &  JLA binned sample  &     BAO sample  & JLA+BAO samples &  JLA & BAO &  JLA+BAO \\
\hline
Linear & $-0.973\pm 0.032$, $0.195\pm 0.685$ & $-0.885\pm0.133$, $1.258\pm 4.139$  
&$-1.015\pm0.024$, $0.271\pm0.616$ &$0.05$ &$0.73$  &$0.86$ \\
CPL & $-0.982\pm 0.045$, $-0.190\pm 1.632$ & $-0.858\pm0.187$, $-1.797\pm 9.441$ 
& $-1.024\pm0.033$, $-0.283\pm1.419$ & $ 0.03$  &$0.73$ &$0.83$\\
BA& $-0.993\pm 0.034$, $-0.068\pm 0.388$ &  $-0.621\pm 0.119$, $ -1.707\pm 2.731$ 
& $-0.892\pm 0.024$, $-0.535\pm 0.450$& $0.02$ &$0.60$ &$0.80$\\
\hline
\hline\\
\end{tabular}\label{tab:full_bestfits}}
\end{minipage}
\end{table*}}}

\subsection{Background analysis: results}

For our background analysis we include the Planck data \cite{Ade:2015xua}, were the selected priors
for $\Omega_m$ and $\Omega_b$ are obtained from a forecast of CMB observations with this astrophysical mission.
In Figure \ref{fig:cosmo_test2} we show the confidence contours for 
the three parameterisations using only the JLA data set. The results 
of the joint SNe Ia + BAO analysis are shown in Figure \ref
{fig:cosmo_test4} where we observe  a clear compatibility with the 
$\Lambda$CDM model.  
To compare the tension \cite{EscamillaRivera:2011qb} among datasets, we compute the so-called $\sigma$-distance, $d_{\sigma}$, between 
the best fit points of each parameterisation and of the $\Lambda$CDM model obtained from the SNe Ia, BAO and the total SNe Ia + BAO analyses.
Following \cite{numerical}, the $\sigma$-distance is calculated by solving 
\begin{equation}
1- \Gamma(1,\vert\Delta \chi_{\sigma}^2/2\vert)/\Gamma(1) = \mathrm{erf}(d_{\sigma}/\sqrt{2}), \label{eq:dsigma}
\end{equation}
where $\Gamma$ and $\mathrm{erf}$ are the Gamma and the error functions, respectively, and $\Delta{\chi_{\sigma}^2}(w_0)=\chi_{\text{Total}}^2 (w_{0_{\text{JLA + BAO}}}) - \chi_{\text{Total}}^{2}(w_{0_{\text{JLA}}})$. We follow the same rule for $w_1$.

The \textit{tension} between probes seems to be reduced when we 
use the BA parameterisation. 
From Table \ref{tab:full_bestfits} we 
also observe that the best fit obtained by using the {BA} 
case is in better agreement with $\Lambda$CDM, around 1\% of 
difference than using the {CPL} case. 

\section{Perturbative Analysis}
\label{sec:linear_perturbations}

Apart from the evolution of the homogeneous part of the dark energy 
parameterisations, the linear perturbations are indeed a substantial 
analysis to understand their evolution. In order to describe their 
dynamics when a scalar field is included, let us write the Einstein 
equations as
\begin{eqnarray}
R_{\mu\nu}-\frac{1}{2}g_{\mu\nu}R = 8\pi 
GT^{b}_{\mu\nu}+\phi_{;\mu}\phi_{;\nu} -\frac{1}{2}g_{\mu\nu} 
\phi_{,a}\phi^{a}_{,} +g_{\mu\nu}V,  \label{eq: einstein_t}
\end{eqnarray}
which preserves the conservation equation 
\begin{eqnarray}
 {T^{\mu\nu}_{b}}_{;\mu}=0, \label{eq:em_eq}
\end{eqnarray}
with the Klein-Gordon equation:
\begin{eqnarray}
\Box\phi =-V_{\phi}, \label{eq:dalambert}
\end{eqnarray}
where $\phi^2=8 \pi G \psi^2$ and $V=8 \pi G U$.
Rewriting Eq.(\ref{eq: einstein_t}) as
\begin{eqnarray}
R_{\mu\nu} = 8\pi G \left(T_{\mu\nu}^{b} -\frac{1}{2} g_{\mu\nu} 
T^{b}\right) +\phi_{,\mu} \phi_{,\nu} -g_{\mu\nu}V,
\end{eqnarray}
we can perturbed to obtain
\begin{eqnarray}
\delta R_{\mu\nu} &=&8\pi G \left(\delta T_{\mu\nu}^{b} -\frac{1}{2} 
h_{\mu\nu}T^{b} -\frac{1}{2} g_{\mu\nu}\delta T^{b}\right) + 
\left(\delta \phi_{,\mu} \phi_{,\nu} +\phi_{,\mu} 
\delta\phi_{,\nu}\right) -h_{\mu\nu} V -g_{\mu\nu} \delta V. \quad
\end{eqnarray}
If we consider the component $\mu =0$, $\nu=0$, we obtain, with the synchronous coordinate condition,
\begin{eqnarray}
\frac{1}{2}\ddot{h} +\left(\frac{\dot{a}}{a}\right) \dot{h} =
4\pi G \,
\delta\rho_m +2\dot{\phi} \, \delta\dot{\phi}  -V_{\phi} \, \delta\phi. 
\label{eq:h}
\end{eqnarray}
The variation of the energy-momentum tensor Eq.(\ref{eq:em_eq}) 
sets the following 
\begin{eqnarray}
\delta \left( T_{b;\mu}^{\mu\nu} \right) =0, \quad \rightarrow \quad 
\dot{\delta}=\frac{\dot{h}}{2}.
\end{eqnarray}
Using this latter and assuming $\delta=\delta\rho_m/\rho_m$, $8 \pi G \rho_m= 
3  H_{0}^{2} \Omega_m /a^3$, $\delta \phi =\nu$
in 
Eq.(\ref{eq:h}) we have
\begin{eqnarray}
\ddot{\delta} +2\left(\frac{\dot{a}}{a}\right)\dot{\delta} 
-\frac{3}{2}H_0^2\Omega_m a^{-3} \delta =2\dot{\phi} \dot{\nu} -V_{\phi} \nu. 
\label{eq:pert1}
\end{eqnarray}
Now, the variation of the D'Alembert operator of the scalar field 
Eq.(\ref{eq:dalambert}) can be computed as
\begin{eqnarray}
\delta(\Box\phi) =-\delta V_{\phi} =-V_{\phi\phi} \nu , 
\label{eq:eq:dalambert2}
\end{eqnarray}
then
\begin{eqnarray}
\delta(\Box\phi) &=& 
\delta\left[g^{\rho\sigma}\left(\phi_{,\rho,\sigma} 
-\Gamma_{\rho\sigma}^{\lambda}\phi_{,\lambda}\right)\right]  
\nonumber\\ &=&-h^{\rho\sigma} \left(\phi_{,\rho,\sigma} 
-\Gamma_{\rho\sigma}^{\lambda}\phi_{,\lambda}\right) + g^{\rho\sigma} 
\left(\delta\phi_{,\rho,\sigma} 
-\Gamma_{\rho\sigma}^{\lambda}\delta\phi_{,\lambda} 
-\chi_{\rho\sigma}^{\lambda}\phi_{,\lambda}\right).
\end{eqnarray}
Using Eq.(\ref{eq:eq:dalambert2}) and the component when $\lambda=0$ we 
finally have
\begin{eqnarray}
\ddot{\nu} +3\left(\frac{\dot{a}}{a}\right)\dot{\nu} 
+\left(\frac{k^2}{a^2}\right)\nu 
-\left(\frac{\dot{h}}{2}\right)\dot{\phi} =-V_{\phi\phi} \nu. 
\label{eq:pert2}
\end{eqnarray}
Also, as we did for Eq.(\ref{eq:pert1}) we can rewrite 
Eq.(\ref{eq:pert2}) as 
\begin{eqnarray}
\ddot{\nu} +3\left(\frac{\dot{a}}{a}\right)\dot{\nu} + 
\left(\frac{k^2}{a^2} +V_{\phi\phi}\right)\nu = 
\dot{\delta}\dot{\phi}.
\end{eqnarray}

Finally, taking $\dot{\delta}=\dot{a}\delta^{\prime}$, where the prime 
denotes derivatives with respect to the scale factor, we can compute 
the following perturbation equations for a scalar field with a specific 
set of $\dot{\phi}$ and $V_{\phi}$
\begin{eqnarray}
&&\delta'' +\left(\frac{2}{a}+\frac{\H'}{\H}\right)\delta' 
-\frac{3}{2}\frac{H_0^2 \Omega_m}{\H^2 a^3 }\delta =2\phi^\prime \nu' 
-\frac{V_{\phi}}{\H^2} \nu, \label{eq:pertb1} \\ && \nu'' 
+\left(\frac{3}{a}+\frac{\H'}{\H}\right)\nu' + \left[\left(\frac{k 
}{a}\right)^2 +V_{\phi\phi}\right] \frac{\nu}{\H^2} =\phi^\prime 
\delta', \quad\quad \label{eq:pertb2} 
\end{eqnarray}
where $\H=\dot{a}$ and $\phi^\prime=\dot{\phi}/\H.$

\subsection{Solutions of linear perturbations and CMB analysis}

\begin{figure}[tbp] 
\centering 
\includegraphics[width=0.8\textwidth,origin=c,angle=0]{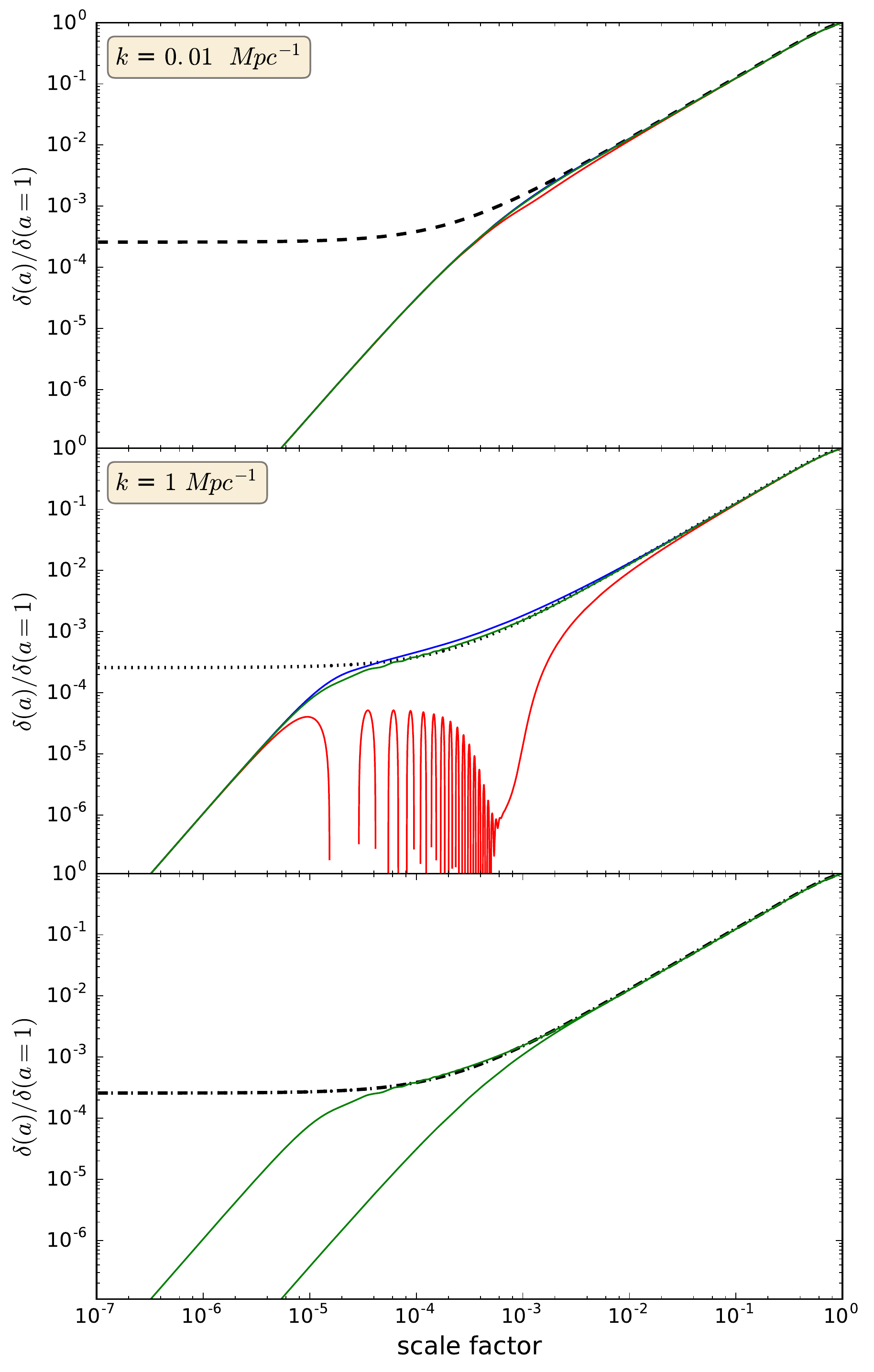} 
\caption{\label{fig:spectrum_k} $k$ modes for the CPL parameterisation. 
Top: The mode $k=0.01 Mpc^{-1}$ of the $\delta$ solution of 
(\ref{eq:pertb2}) in black-dashed, compared with the output of the modified 
CAMB for cold dark matter ($\delta_c$ in blue), baryons ($\delta_b$ in red) and all matter ($\delta_m$ in green).
Middle: The same for the mode $k=1$ ($\delta$ in black-dotted). Bottom: The two $\delta$ modes are undistinguishable
and overlap to the CAMB solutions of $\delta_m$ after the recombination era.
For brevity, we only show the behaviour for CPL model 
since the results for the other two parameterisations are essentially the same.}

\label{fig:kspectrum}
\end{figure}
In the attempt to account for a complete treatment of the linear 
evolution of perturbations from the entry 
in the horizon of perturbations produced by inflation until today 
for each scale, we modified the Boltzmann code named CAMB \footnote{ 
\href{http://camb.info} {http://camb.info}} \cite{camb}, that solves 
numerically the fluid equations following \cite{wellerlewis}:

\begin{eqnarray} 
\delta_i' + 3\H(\hat{c}_{s,i}^2-w_i)(\delta_i 
+3\H(1+w_i)v_i/k)+(1+w_i)kv_i + 3\H w'_i v_i / k = -3(1+w_i)h' \label{eqn:di1} \\
 v_i' + \H(1-3\hat{c}_{s,i}^2)v_i= k \hat{c}_{s,i}^2 \
 \delta_i/(1+w_i) \, 
\label{eqn:di2} \end{eqnarray} 
where derivatives are respect the conformal time, $\H$ is the conformal Hubble 
parameter, $v_i$ is the velocity, $w_i\equiv p_i/\rho_i$, $h' = (\delta 
a/a)'$, and $\hat{c}_s^2$ is the sound speed evaluated in the frame co-moving with dark energy 
($\hat{c_s}=1$ for quintessence).

In order to avoid the crossing instability problem at the phantom divide  
line, i.e. $w=-1$, the CAMB code provides a module that implements 
the Parameterized Post-Friedmann (PPF) approach \cite{PPF} for the 
CPL model. We modified it to also include the Linear and BA 
parameterisations. The advantage of this approach is to replace the 
condition on the dark energy pressure perturbation with a 
relationship between the momentum density of this dark component and 
that of the other components on the large scales, providing a 
well-controlled approximation for any model where the energy and 
momentum of the dark energy are separately conserved.

The solution of (\ref{eq:pert1}) and (\ref{eq:pert2}) are compared with 
the solution of (\ref{eq:pertb1}) and (\ref{eq:pertb2}) for different 
modes, $k=0.01 \,  h$Mpc$^{-1}$ and $k=1 \,  h$Mpc$^{-1}$, and shown in 
Figure~\ref{fig:kspectrum} for the CPL case. We start the computation 
of (\ref{eq:pert1}) and (\ref{eq:pert2}) inside the radiation era at 
$a=10^{-7}$ with $\dot{\delta}=\delta \dot{a} / (a+a_{eq}/1.5)$, where 
$a_{eq}$ is the scale factor at radiation-matter equivalence 
\cite{Peebles_book}, and, concordantly, we consider the radiation term 
contribution in the Friedmann equation. In this figure  we see that the 
evolution of $\delta(a)$ is scale invariant for the parametrisations 
here considered. Qualitatively, the scalar field undergo damped 
oscillations for scales $k>a^2 V_{\phi \phi}$. On these scales the 
scalar field will not contribute to the total gravitational potential 
and can be approximated as homogeneous. The difference of the solutions 
at early time, occurs because CAMB takes into account the entrance of 
the perturbations at the horizon scale. On the contrary, for eqs 
(\ref{eq:pert1}) and (\ref{eq:pert2}), we set the initial conditions 
immediately inside the horizon scale at $a=10^{-7}$, when, actually, 
the perturbations at scales $k=1 \,  h$Mpc$^{-1}$ and $k=0.01 \, 
h$Mpc$^{-1}$ have not entered inside the horizon yet. Accordingly, 
Figure~\ref{fig:kspectrum} shows that the solutions overlap after the 
perturbations have entered in the horizon (in CAMB), and not before.

\begin{figure}[tbp]
\centering
\includegraphics[width=0.8\textwidth,origin=c,angle=0]{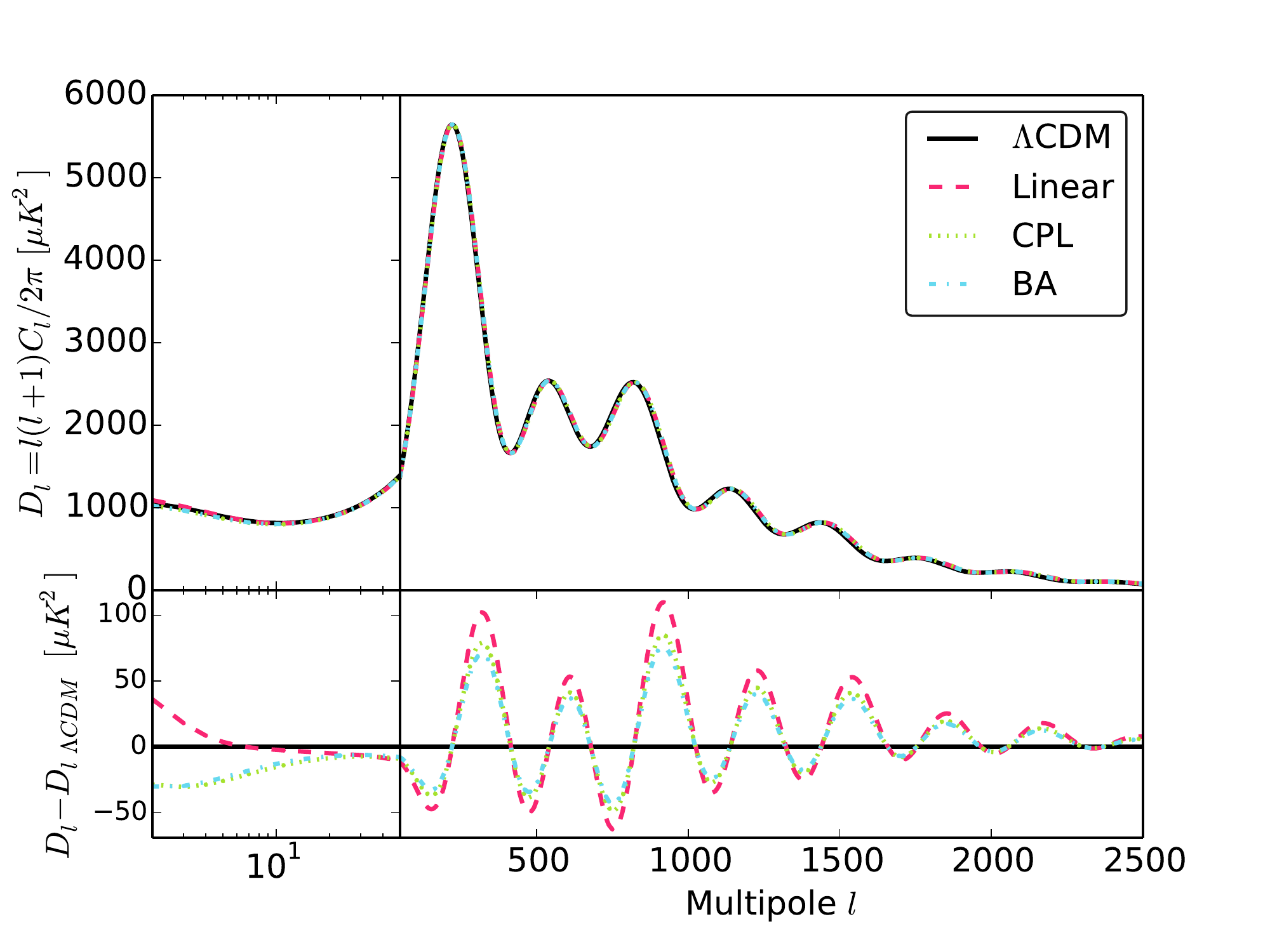}
\includegraphics[width=0.8\textwidth,origin=c,angle=0]{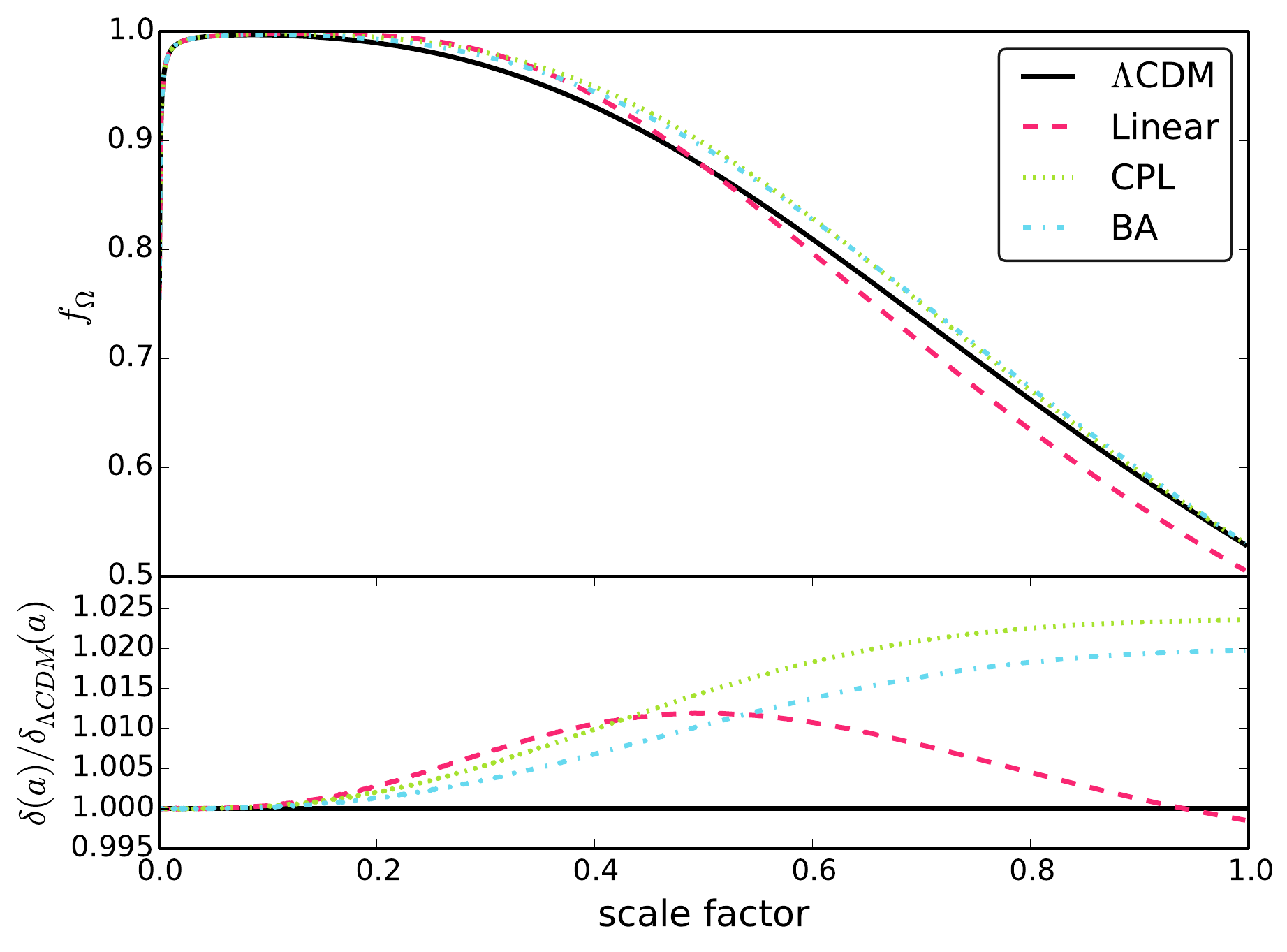}
\caption{\label{fig:power_spectrum}
Top: Linear analysis
using PPF. The CMB $C^{TT}_{l}$ power spectrum versus 
multipole moment $l$ using the best fit values obtained for each dark 
energy parameterisation using the combined JLA+BAO data set. 
Bottom: The evolution of $f_\Omega$ and $\delta(a)$ for each model.
}
\end{figure}

The comparison of the results obtained from CAMB for the different dark 
energy parametrizations are shown in Figure \ref {fig:power_spectrum}. 
The CMB power spectra of our models with respect to the $\Lambda$CDM 
cosmology are shown in the upper panel. All the models share the same 
$\log$ power of the primordial curvature perturbations 
ln$(10^{10}A_s)=3.09$ and the same scalar spectrum index $n_s=0.966$, 
with $k_0=0.05$ Mpc$^{-1}$ as indicated by \cite {Ade:2015xua}. Fixing 
the set of parameters by the SNe Ia constraints for each model, the 
discrepancies from the $\Lambda$CDM scenario favor the BA 
parameterisation. A finest future analysis could also consider the 
employment of the code COSMOMC \footnote {\href 
{http://cosmologist.info/cosmomc}{http://cosmologist.info/cosmomc}} 
\cite{cosmomc} that provides a joint analysis of background and 
perturbation exploring a large dimensional space of parameters. 

Finally, as mentioned earlier, the $\delta(a)$ evolution is 
scale-invariant. Therefore, we can have a complete description of the 
growth rate of structure,  defined as 
$f_\Omega=\frac{d\ln{\delta}}{d\ln{a}}$, by considering a single mode. 
In the bottom panel of Figure \ref{fig:power_spectrum}, it is shown 
that the variations with respect to the different models do not exceed 
$\sim 2.5\%$ at $z=0$.

\section{Non linear power spectrum} 
\label{sec:non_linear}

In this section we estimate the non-linear contribution of the $w(z)$ parameterisations to the matter power spectrum. The aim here is to detect any observational signature that may distinguish among these models.
Such contribution can be  derived in a straightforward way by performing time expansive numerical simulations. However, this is not 
necessary. A common approach in the prediction of the $P(k)$ is to compute
the non linearities using the Halofit recipe \cite{halofit1} revised in \cite {halofit2}. In \cite{hmcode}, it was 
introduced the HMcode\footnote{\href 
{https://github.com/alexander-mead/HMcode}{https://github.com/alexander-mead/HMcode}}, a fit of the Halo model \cite{halomodel} on the 
Coyote suite simulations \cite{coyote} including also the effect of 
baryons physics as quantified in \cite{agnsim}, that implemented 
star formation, supernovae and active galactic nucleus feedback. 
Despite these methods are built on simulations of models including 
$w=const.$ only, they are employed for $w=w(a)$ models \cite 
{Ade:2015xua}, by entering an evident bias in the forecasts.

\begin{figure}[tbp] \centering \includegraphics 
[width=0.8\textwidth,origin=c,angle=0]{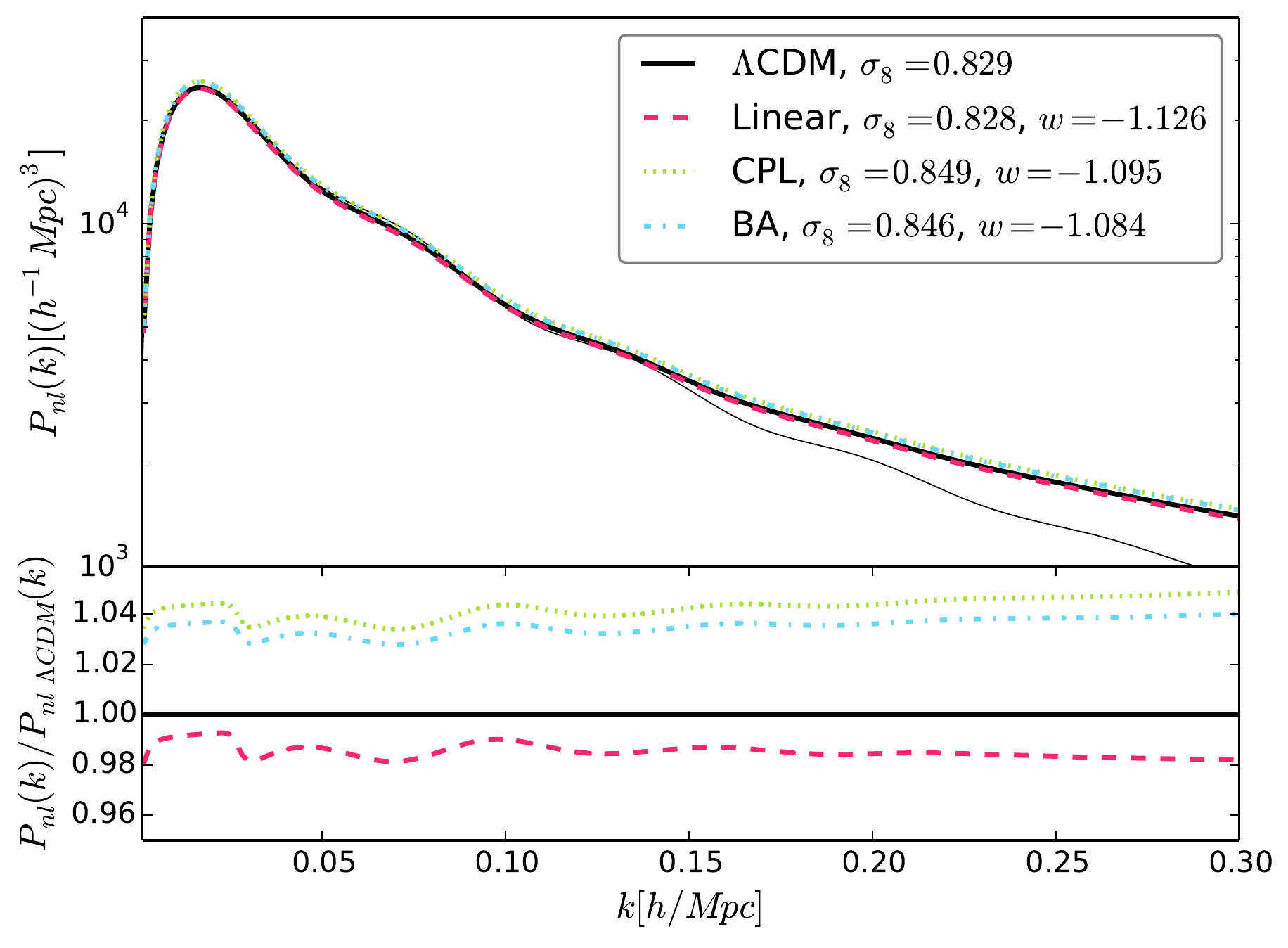} \caption{\label{fig:power_spectrum2} 
Non linear power spectrum derived using the PKequal package, as discussed in the text. 
$\sigma_8$ values are computed on the linear spectra and the w values are the equivalent w-constant models that reproduce the non linear spectra for every model, following \cite{2009JCAP...03..014C, pkequal}. The thin line represents the linear matter power spectrum predicted by the $\Lambda$CDM 
model. }
\end {figure}

In Ref.~\cite{2009JCAP...03..014C}, it was found a  spectral equivalence 
between $w=w(a)$ models and $w=const$, that drastically simplifies 
the task to find a universal Halo model. Previous authors \cite 
{2007MNRAS.380.1079F} had shown how spectral predictions for 
constant--$w$ models at $z=0$ can also be used to fit spectra of CPL 
cosmologies with a precision $\sim 1\%$. To meet such precision at 
non-zero redshift, a new technique is needed, that it was defined in 
\cite{2009JCAP...03..014C} and tested for several models \cite 
{2009JCAP...03..014C,2010JCAP...08..005C}, both through purely 
gravitational simulations and through simulations including baryon 
physics \cite{2011MNRAS.412..911C}.  Recently, in \cite{pkequal} it 
was described a public package called PKequal\footnote{\href 
{https://github.com/luciano-casarini/PKequal}{https://github.com/luciano-casarini/PKequal}}, that extends both Halofit and Coyote suite 
from $w=const$ to CPL models. In this work, we modified the 
PKequal code in order to also include the Linear and BA cases and chose  to extend the predictions of the code described in \cite
{coyote}. The results are shown in Figure \ref 
{fig:power_spectrum2}. We observe that most of the discrepancies is due to 
the linear evolution, while the non-linear contribution is almost 
the same for all the scales. Indeed the $P(k)$ differs between the 
models according to the $\delta(a)$ evolution at $z=0$ (see Figure 
\ref{fig:power_spectrum2}). Clearly, the current uncertainties on the cosmological parameters do  
not allow to discriminate between the parameterisation models. It is expected that future weak lensing 
surveys that aim to distinguish matter distribution at $1\%$ level, 
e.g. Euclid \cite{euclid_data}, will be able to discriminate time-dependent dark energy parameterisations from the standard $\Lambda$CDM model using the observables discussed in this analysis.  

\section{Conclusions}
\label{sec:conclusions}

We have studied three dark energy parameterisations 
with two free parameters (Linear, CPL and BA), which make suitable 
for fitting the current JLA, BAO and the combination of these 
datasets. By using these models, it was possible to set the dynamics 
of the dark energy in a scalar field representation. In Table \ref
{tab:full_bestfits} it was reported the best-fit parameters for each model 
using the current cosmological data. Also, it is important to remark 
that we have presented a complete treatment of the linear evolution of 
perturbations from the entry of 
perturbations produced by inflation in the horizon until today. We have calculated the numerical perturbations for each parametrisation  and shown that the differences of the growth factors for 
extreme values of $k$ are almost negligible. Furthermore, using a suitable spectral equivalence, we have shown that 
the non-linear contribution of the selected dark energy 
parameterisations to the matter power spectra is almost the same for 
all scales of interest. Finally, we have 
observed that, for the current uncertainties on the cosmological parameters, the power spectrum for our three selected 
parameterisations is indistinguishable from the one predicted by the 
standard model. We expect the next generation of cosmology experiments to be able
to distinguish between time-dependent $w(z)$ models and the standard $\Lambda$CDM cosmology using the observables discussed in the present analysis.

\acknowledgments

C.E-R is supported by CNPq Brazil Project 502454/2014-8. L. C. thanks to CAPES. J.C.F thanks to CNPq and FAPES. 
J.S.A. thanks CNPq, FAPERJ and INEspa\c{c}o for the financial support.



\end{document}